# The Non-Euclidean Hydrodynamic Klein–Gordon Equation with Perturbative Self-Interacting Field

**Piero Chiarelli** [1,2]

[1] National Council of Research of Italy, Area of Pisa, Moruzzi 1, 56124 Pisa, Italy; pchiare@ifc.cnr.it; Tel.: +39-050-315-2359; Fax: +39-050-315-2166

[2] Interdepartmental Center "E.Piaggio", University of Pisa, 56124 Pisa, Italy

**Abstract:** In this paper the quantum hydrodynamic approach for the Klein–Gordon equation (KGE) owning a perturbative self-interaction term is developed. The generalized model to non-Euclidean space–time allows for the determination of the quantum energy impulse tensor density of mesons, for the gravitational equation of quantum mechanical systems.

**Keywords:** quantum hydrodynamic representation; Bhom–Madelung approach; self-interaction; Einstein equation for quantum mechanical system

## 1. Introduction

A longstanding objective of modern physics is to describe the quantum mechanics and the quantum fields in the framework of general relativity. On an ordinary scale, the two theories are decoupled and even if theoretical models are available [1–16], it not easy to find experimental confirmations.

A first physical evidence of quantum gravity (QG) comes by way of the corrections to the Newton law, derived from the Galilean limit of the theory [17–29]. The Hawking radiation of a black hole can also be justified in the frame of QG [30–38].

Moreover, given that the uncertainty principle forbids a black hole (BH) collapse [39–44] and that the quantization of a black hole requires the existence of a fundamental state with a minimum mass, the BH cannot have a mass smaller than that of Planck and comparable with those of elementary particles [45].

Another phenomenon connected to QG is the cosmological constant that it is needed to model the motion of galaxies on the universal scale [46–54].

If we try to define the QG equation from the minimum action principle (MAP), the quantizing action is not self-contained under renormalization, since loop diagrams generate terms not present in the initial expression [55–59].

In the present work the author makes use of the hydrodynamic quantum description (HQD) [60–68] that has the advantage of being embedded in a classical-like formalism, where the quantization is implicitly contained into the equation of evolution [68], for deriving a gravity equation for quantum mechanical systems.

The quantum hydrodynamic approach, firstly developed by Madelung [62], describes the evolution of the complex wavefunction $\Psi = |\Psi| exp \frac{i}{\hbar} S$ as a function of the two real variables, $|\Psi|$ and $S$ [60,62,63,69–73]. The model gives rise to classical-like analogy describing the motion of particles' density $n = |\Psi|^2$ owning the 4-impulse $p_\mu = -\partial_\mu S$.

As shown by Weiner et al. [71], the outputs of the quantum hydrodynamic model agree with the outputs of the Schrödinger problem, but not only for the semi-classical limit or for a single particle [62,63,69]. More recently, Koide and Kodama [72], showed that it agrees with the outputs of the stochastic variational method.

Recently, the author of this article has shown that the hydrodynamic approach can be derived from the properties of vacuum on a small scale [73].

Moreover, as shown by Bohm and Hiley [74,75] the hydrodynamic approach can be generalized for the description of the quantum fields.

The present work develops the quantum hydrodynamic form of the Klein–Gordon equation (KGE) containing an additional perturbative self-interaction term.

The interest of such a description lies in the fact that this kind of KGE can describe the states of bosons, such as mesons. The goal of this paper is to obtain the energy–impulse tensor of such a particle, for the coupling with the Einstein gravitational equation (GE) for quantum mechanical systems [68].

As shown in study [68], the quantum mechanical QG defines the gravitational coupling of classical fields and the commutation rules in curved space to be used for their quantization.

The paper is organized as follows: in Section 2, the hydrodynamic KGE with a perturbative self-interaction term is derived for scalar particle and for a charged boson. In Section 3, the formulae are generalized to a non-Euclidean space–time.

## 2. The Hydrodynamic KGE with Perturbative Self-Interaction

In this section, the Euclidean hydrodynamic representation of the KGE is derived for a scalar uncharged boson with a perturbative self-interaction term that reads

$$\partial_\mu \partial^\mu \psi = -\frac{m^2 c^2}{\hbar^2} \psi + \lambda |\psi|^2 \psi \tag{1}$$

Following the procedure given in studies [45,68] the hydrodynamic equations of motion are given by the Hamilton–Jacobi type equation

$$g^{\mu\epsilon} \partial_\mu S \, \partial_\epsilon S - \hbar^2 \left( \frac{\partial_\mu \partial^\mu |\psi|}{|\psi|} - \lambda |\psi|^2 \right) - m^2 c^2 = 0 \tag{2}$$

coupled with the current equation [67,69]

$$\partial_\mu \left( |\psi|^2 \partial^\mu S \right) = 0 \tag{3}$$

where

$$S = \frac{\hbar}{2i} ln[\frac{\psi}{\psi^*}] \tag{4}$$

and where

$$\frac{|\psi|^2}{m} \partial_\mu S = J_\mu = \frac{i\hbar}{2m} (\psi^* \partial_\mu \psi - \psi \partial_\mu \psi^*) \tag{5}$$

where the signature $(1, -1, -1, -1)$ has been used.

Moreover, being the 4-impulse of the hydrodynamic analogy

$$p_\mu = -\partial_\mu S \tag{6}$$

it follows that

$$J_\mu = (c\rho, -J_i) = -|\psi|^2 \frac{p_\mu}{m} \tag{7}$$

where

$$\ldots = \frac{|\OE|^2}{mc^2}\frac{\partial S}{\partial t} \qquad (8)$$

Moreover, by using (6), Equation (2) can be rewritten as

$$\partial_{\_}S\partial^{\sim}S = p_{\_}p^{\sim} = \left(\frac{E^2}{c^2} - p^2\right) = m^2c^2\left(1 - \frac{V_{qu}}{mc^2}\right) \qquad (9)$$

where

$$V_{qu} = -\frac{\hbar^2}{m}\left(\frac{\partial_{\_}\partial^{\sim}|\OE|}{|\OE|} - \} |\OE|^2\right), \qquad (10)$$

and where $p^2 = p_i p_i$ is the modulus of the spatial 3-impulse.

As shown in studies [67,68], given the hydrodynamic Lagrangean function

$$L = \frac{dS}{dt} = \frac{\partial S}{\partial t} + \frac{\partial S}{\partial q_i}\dot{q}_i = -p_{\_}\dot{q}^{\sim}$$

$$= \frac{1}{2}\frac{\sum_n a_n |\OE_n| exp[\frac{iS_n}{\hbar}]\left(\frac{\hbar}{i}\dot{q}^{\sim}\partial_{\_} \ln|\OE_n| + L_n\right)}{\sum_n a_n |\OE_n| exp[\frac{iS_n}{\hbar}]} \qquad (11)$$

$$-\frac{1}{2}\frac{\sum_n a^*_n |\OE_n| exp[\frac{-iS_n}{\hbar}]\left(\frac{\hbar}{i}\dot{q}^{\sim}\partial_{\_} \ln|\OE_n| - L_n\right)}{\sum_n a^*_n |\OE_n| exp[\frac{-iS_n}{\hbar}]}$$

Equation (2) can be expressed by the following system of Lagrangean equations of motion

$$p_{\_} = -\frac{\partial L}{\partial \dot{q}^{\sim}}, \qquad (12)$$

$$\dot{p}_{\_} = -\frac{\partial L}{\partial q^{\sim}}. \qquad (13)$$

that for the eigenstates read

$$p_{n\_} = -\frac{\partial L_n}{\partial \dot{q}^{\sim}}, \qquad (14)$$

$$\dot{p}_{n\_} = -\frac{\partial L_n}{\partial q^{\sim}} \qquad (15)$$

where the Lagrangean for positive and negative-energy states $L_{+n}$ and $L_{-n}$, respectively reads

$$L_{\pm n} = (\pm)L_n = (\pm)-\frac{mc^2}{\mathsf{x}}\sqrt{1 - \frac{V_{qu(n)}}{mc^2}} = (\pm)-\frac{mc^2}{\mathsf{x}}\sqrt{1 + \frac{\hbar^2}{m^2c^2}\left(\frac{\partial_{\_}\partial^{\sim}|\OE_n|}{|\OE_n|} - \}|\OE\right.} \qquad (16)$$

where $c$ is the speed of light and $\chi = \dfrac{1}{\sqrt{1-\dfrac{\dot{q}^2}{c^2}}}$. Generally speaking, for eigenstates, for which it holds $E = E_n = const$, it follows that

$$\left(\frac{E_n^2}{c^2} - p_n^2\right) = m^2 c^2 \left(1 - \frac{V_{qu(n)}}{mc^2}\right)$$
$$= m^2 \chi^2 c^2 \left(1 - \frac{V_{qu(n)}}{mc^2}\right) - m^2 \chi^2 \dot{q}^2 \left(1 - \frac{V_{qu(n)}}{mc^2}\right) \tag{17}$$

from where it follows that

$$E_n = \pm m\chi c^2 \sqrt{1 - \frac{V_{qu(n)}}{mc^2}} = \pm m\chi c^2 \sqrt{1 + \frac{\hbar^2}{m^2 c^2}\left(\frac{\partial_\sim \partial^\sim |\Phi_n|}{|\Phi_n|} - \right)/|\Phi_n|^2} \tag{18}$$

(where the minus sign stands for antiparticles) and, by using (17), that

$$p_{n\sim} = \pm m\chi \dot{q}_\sim \sqrt{1 - \frac{V_{qu(n)}}{mc^2}} = \frac{E_n}{c^2}\dot{q}_\sim. \tag{19}$$

Following the hydrodynamic protocol [68], the eigenstates are represented by the stationary solutions of the hydrodynamic equations of motion obtained by deriving $p_{\sim(\dot{q},q)}$ from (14) and then inserting it into (15) that leads to

$$\frac{dp_{n\sim}}{ds} = -\frac{\chi}{c}\frac{\partial L_n}{\partial q^\sim} = \pm \frac{d}{ds}\left(mcu_\sim \left(\sqrt{1 - \frac{V_{qu)(n)}}{mc^2}}\right)\right)$$
$$= \pm mc \frac{\partial}{\partial q^\sim}\sqrt{1 - \frac{V_{qu(n)}}{mc^2}} \tag{20}$$

where

$$u_\sim = \frac{\chi}{c}\dot{q}_\sim, \tag{21}$$

and to

$$\pm mc\sqrt{1 - \frac{V_{qu(n)}}{mc^2}}\frac{du_\sim}{ds} = (\pm) - mcu_\sim \frac{d}{ds}\left(\sqrt{1 - \frac{V_{qu(n)}}{mc^2}}\right) \pm mc\frac{\partial}{\partial q^\sim}\left(\sqrt{1 - \frac{V_{qu(n)}}{mc^2}}\right) = -\frac{\chi}{c}\frac{\partial T_{\pm n\sim}^\epsilon}{\partial q^\epsilon} \tag{22}$$

where for eigenstates, the quantum energy-impulse tensor (QEIT) $T_{n\sim}^\epsilon$ reads [45,68]

$$T_{\pm n\sim}^\epsilon = \pm T_{n\sim}^\epsilon = \pm\left(\dot{q}_\sim \frac{\partial L_n}{\partial \dot{q}_\epsilon} - L_n u_\sim^\epsilon\right) = \pm \frac{mc^2}{\chi}\sqrt{1 - \frac{V_{qu(n)}}{mc^2}}\left(u_\sim u^\epsilon - u_\sim^\epsilon\right). \tag{23}$$

leading to the quantum energy impulse tensor density (QEITD) [45,68]

$$T_{n\sim}^\epsilon = \dot{q}_\sim \frac{\partial L_{nn}}{\partial \dot{q}_\epsilon} - L_{nn}u_\sim^\epsilon = |\Phi_n|^2 \left(\dot{q}_\sim \frac{\partial L_n}{\partial \dot{q}_\epsilon} - L_n u_\sim^\epsilon\right) = |\Phi_n|^2 T_{n\sim}^\epsilon \tag{24}$$

where $L = |Œ|^2 \mathcal{L}$ is the (hydrodynamic) Lagrangian density and $\mathcal{L}$ is the hydrodynamic Lagrangian function.

The quantization condition is brought inside the above quantum hydrodynamic relations by the quantum potential $V_{qu(n)}$ contained in the term $\sqrt{1 - \frac{V_{qu(n)}}{mc^2}}$. The analog quantities of the classical limit are recovered for $\hbar \to 0$, and the energy–impulse tensor (23) in Equation (22) leads to the classical equation of motion.

Moreover, by using the identity

$$S_n = \frac{\hbar}{2i} ln[\frac{Œ_n}{Œ_n*}], \qquad (25)$$

the QEITD (24) can be written as a function of the wave function as following

$$T_{n\sim}{}^{\epsilon} = \pm |Œ|^2 c^2 \left(\frac{\partial S_n}{\partial t}\right)^{-1} \left(p_\sim p^\epsilon - p_r p^r u_\sim{}^\epsilon\right)$$

$$= \pm m |Œ_n|^2 c^2 \left(\frac{\frac{\hbar}{2im^2 c^2} \partial ln[\frac{Œ_n}{Œ*_n}]}{\partial t}\right)^{-1} \qquad (26)$$

$$\left(\left(\frac{\hbar}{2mc}\right)^2 \frac{\partial ln[\frac{Œ_n}{Œ*_n}]}{\partial q^\sim} \frac{\partial ln[\frac{Œ_n}{Œ*_n}]}{\partial q_\epsilon} + \left(1 - \frac{V_{qu(n)}}{mc^2}\right) u_\sim{}^\epsilon\right)$$

*Charged Boson*[M1]

In the case of a charged boson, Equations (1)–(3) read, respectively,

$$D_\sim D^\sim Œ = -Œ \left(\frac{m^2 c^2}{\hbar^2} - \} |Œ|^2\right) \qquad (27)$$

where $D_\sim = \partial_\sim - \frac{e}{i\hbar} A_\sim$,

$$\left(\partial_\sim S + eA_\sim\right)\left(\partial^\sim S + eA^\sim\right) = m^2 c^2 + \hbar^2 \left(\frac{\partial_\sim \partial^\sim |Œ|}{|Œ|} - \} |Œ|^2\right) \qquad (28)$$

$$\frac{\partial J_\sim}{\partial q_\sim} = 0 \qquad (29)$$

where the 4-current $J_\sim$ reads

$$J_\sim = (c..., -J_i) = \frac{\hbar}{2im}(Œ * \left(\partial_\sim - \frac{e}{i\hbar} A_\sim\right) Œ - Œ \left(\partial_\sim + \frac{e}{i\hbar} A_\sim\right) Œ*)$$

$$= -\frac{|Œ|^2}{m}\left[p_\sim - eA_\sim\right] = -\frac{|Œ|^2}{m} f_\sim \qquad (30)$$

and

$$-\partial_\sim S = p_\sim = f_\sim + eA_\sim = (\frac{E}{c}, -p_i) \tag{31}$$

(where $f_\sim$ is the mechanical momentum) [67,68] and where

$$\ldots = -\frac{|Œ|^2}{mc^2}\left[\frac{\partial S}{\partial t} + eW\right]. \tag{32}$$

Moreover, analogously to (9) and (17)–(19), from (28) it follows that

$$f_{n\sim} = \frac{1}{c^2}\left[-\frac{\partial S_n}{\partial t} - eW\right]\dot{q}_\sim$$
$$= \frac{E_n - eW}{c^2}\dot{q}_\sim = p_{n\sim} - eA_\sim \tag{33}$$

that leads to

$$E_n - eW = \pm mx\, c^2\sqrt{1 - \frac{V_{qu(n)}}{mc^2}} = \pm mx\, c^2\sqrt{1 + \frac{\hbar^2}{m^2 c^2}\left(\frac{\partial_\sim \partial^\sim |Œ_n|}{|Œ_n|} - \} |Œ_n|^2\right)}, \tag{34}$$

to

$$p_{n\sim} - eA_\sim = \pm mx\, \dot{q}_\sim\sqrt{1 - \frac{V_{qu(n)}}{mc^2}} = \frac{E_n}{c^2}\dot{q}_\sim \tag{35}$$

and to

$$L_n = \frac{dS_n}{dt} = \frac{\partial S_n}{\partial t} + \frac{\partial S_n}{\partial q_i}\dot{q}_i = -p_{n\sim}\dot{q}^\sim = -\left(\frac{1}{c^2}\frac{\partial S_n}{\partial t} + \frac{e}{c^2}W\right)^{-1} p_{n\sim}(p_n^\sim - eA^\sim)$$
$$= \left(-\frac{mc^2}{x}\sqrt{1 + \frac{\hbar^2}{m^2 c^2}\left(\frac{\partial_\sim \partial^\sim |Œ_n|}{|Œ_n|} - \} |Œ_n|^2\right)} - eA_\sim \dot{q}^\sim\right) \tag{36}$$

that by using (25), as a function of $Œ$ and $A^\sim$, reads

$$L_n = -\frac{i\hbar}{2}c^2\left(\frac{\partial \ln[\frac{Œ_n}{Œ_n^*}]}{\partial t} + \frac{2ie}{\hbar}W\right)^{-1}\frac{\partial \ln[\frac{Œ_n}{Œ_n^*}]}{\partial q^\sim}\left(\frac{\partial \ln[\frac{Œ_n}{Œ_n^*}]}{\partial q_\sim} - \frac{2ie}{\hbar}A^\sim\right). \tag{37}$$

For $\hbar = 0$ the Lagrangean (37) acquires the known classical expression $L = -\left(\frac{mc^2}{x} + eA_\sim \dot{q}^\sim\right)$.

Moreover, with the help of (25), (30), (33)–(35) it follows that

$$T_{n\sim}^\epsilon = -\frac{m|Œ_n|^2 c^2}{x}\sqrt{1 - \frac{V_{qu(n)}}{mc^2}}\left(\left(u_\sim + \left(\sqrt{1 - \frac{V_{qu(n)}}{mc^2}}\right)^{-1}\frac{e}{mc}A_\sim\right)u^\epsilon\right.$$
$$\left. - \left(1 + \left(\sqrt{1 - \frac{V_{qu(n)}}{mc^2}}\right)^{-1}\frac{e}{mc}A_r u^r\right)u_\sim^\epsilon\right) \tag{38}$$

that by using (25), (30) and (35) we can be expressed as a function of the wave function as

$$T_{\sim}^{\epsilon} = |Œ_n|^2 \frac{\hbar c^2}{2i} \left( \frac{\partial}{\partial t} ln[\frac{Œ_n}{Œ_n^*}] - \frac{2ie}{\hbar} W \right)^{-1} \left( \begin{pmatrix} \frac{\partial ln[\frac{Œ_n}{Œ_n^*}]}{\partial q^{\sim}} - \frac{2ie}{\hbar} A_{\sim} \end{pmatrix} \frac{\partial ln[\frac{Œ_n}{Œ_n^*}]}{\partial q_{\epsilon}} \\ - \begin{pmatrix} \frac{\partial ln[\frac{Œ_n}{Œ_n^*}]}{\partial q^r} - \frac{2ie}{\hbar} A_r \end{pmatrix} \frac{\partial ln[\frac{Œ_n}{Œ_n^*}]}{\partial q_r} u_{\sim}^{\epsilon} \right) \quad (39)$$

The above equations are coupled to the Maxwell equation:

$$F^{\sim \epsilon}{}_{,\epsilon} = -4f \, J^{\sim} \quad (40)$$

where [75]

$$F_{\sim \epsilon} = \left( A_{\epsilon, \sim} - A_{\sim, \epsilon} \right) = \left( \partial_{\sim} A_{\epsilon} - \partial_{\epsilon} A_{\sim} \right), \quad (41)$$

and where

$$A_{\sim} = (\frac{W}{c}, -A_i) \quad (42)$$

is the potential 4-vector.

### 3. Non-Euclidean Space–Time

The quartic self-interaction is introduced in the KGE in order to describe the states of charged ($\pm 1$) bosons (e.g., mesons) [76]. The importance of having the hydrodynamic description of bosons [68] lies in the fact that it becomes possible to derive its quantum energy–impulse tensor that can couple them to the Einstein gravitational equation for quantum mechanical systems [68].

By using the General Physics Covariance postulate [3,68], it is possible to derive the non-Euclidean expression of the hydrodynamic model of the KGE:

$$\left( \partial^{\sim} Œ \right)_{,\sim} = \frac{1}{\sqrt{-g}} \partial_{\sim} \left( \sqrt{-g} \, g^{\sim \epsilon} \partial_{\epsilon} Œ \right) = -Œ \left( \frac{m^2 c^2}{\hbar^2} - \} |Œ|^2 \right) \quad (43)$$

Equations (2) and (3) in a non-Euclidean space–time read, respectively,

$$g^{\sim \epsilon} \partial_{\sim} S \, \partial_{\epsilon} S + mV_{qu} - m^2 c^2 = 0$$

$$\frac{1}{\sqrt{-g}} \partial_{\sim} \sqrt{-g} \left( g^{\sim \epsilon} |Œ|^2 \partial_{\epsilon} S \right) = 0 \quad (44)$$

where

$$V_{qu} = -\frac{\hbar^2}{m} \left( \frac{1}{|Œ| \sqrt{-g}} \partial_{\sim} \sqrt{-g} \left( g^{\sim \epsilon} \partial_{\epsilon} |Œ| \right) - \} |Œ|^2 \right). \quad (45)$$

Moreover, by using the Lagrangean function

$$L_n = \frac{dS_n}{dt} = -g_{\sim \epsilon} \, p_n^{\epsilon} \dot{q}^{\sim}, \quad (46)$$

the covariant form of the motion Equations (14) and (15) reads

$$p_{n_\sim} = -\frac{\partial L_n}{\partial \dot{q}^\sim} \qquad (47)$$

$$D_t p_{n_\sim} = -\frac{\partial L_n}{\partial q^\sim}, \qquad (48)$$

where

$$D_t p_\sim = \dot{p}_\sim - \Gamma^\epsilon_{\sim\}} p_\epsilon \dot{q}^\} \qquad (49)$$

is the total covariant derivative respect the time and where $\Gamma^\epsilon_{\sim\}}$ are the Christoffel symbols.

Equations (47) and (48) leads to the motion equation

$$D_t \left( -\frac{\partial L_n}{\partial \dot{q}^\sim} \right) = -\frac{\partial L_n}{\partial q^\sim} \qquad (50)$$

where $L_n$ reads

$$L_n = -\frac{mc^2}{\chi}\sqrt{1-\frac{V_{qu(n)}}{mc^2}} = -mc^2 \sqrt{\frac{g_{\sim\epsilon}\dot{q}^\epsilon \dot{q}^\sim}{c^2}}\sqrt{1-\frac{V_{qu(n)}}{mc^2}}. \qquad (51)$$

From (50) it follows that the motion equation, describing the evolution of the particle density $n = |Œ_n|^2$ moving with the hydrodynamic impulse (6), reads

$$\frac{du_\sim}{dt} - \frac{1}{2}\frac{c}{\chi}\frac{\partial g_{\}|}}{\partial q^\sim}u^\}u^| = \frac{c}{\chi}\frac{\partial \ln\sqrt{1-\frac{V_{qu(n)}}{mc^2}}}{\partial q^\sim} + \frac{c}{\chi}\frac{\partial \ln\chi}{\partial q^\sim} - u_\sim \frac{d}{dt}\ln\sqrt{1-\frac{V_{qu}}{mc^2}} \qquad (52)$$

where the stationary condition $\frac{du_\sim}{dt}=0$, that determines the balance between the "force" of gravity and that one of the quantum potential, defines the stationary equation for the eigenstates

$$\frac{1}{2}\frac{c}{\chi}\partial_\sim g_{\}|}u^\}u^| + \frac{c}{\chi}\partial_\sim \ln\chi = -\frac{c}{\chi}\partial_\sim \ln\sqrt{1-\frac{V_{qu(n)}}{mc^2}} + u_\sim \frac{d}{dt}\ln\sqrt{1-\frac{V_{qu}}{mc^2}}. \qquad (53)$$

where $\frac{1}{g} = |g_{\epsilon\sim}| = -J_{ac}^2$, where $J_{ac}$ is Jacobean of the transformation of the Galilean co-ordinates to non-Euclidean ones and where $g_{\epsilon\sim}$ is the metric tensor defined by the gravitational equation for quantum mechanical systems [68,75]

$$R_{\sim\epsilon} - \frac{1}{2}g_{\sim\epsilon}R_r^{\ r} - \Lambda g_{\sim\epsilon} = \frac{8fG}{c^4}T_{\sim\epsilon} \qquad (54)$$

where the quantum energy impulse tensor density reads

$$T_{(n)\sim\epsilon} = |Œ_n|^2 mc^2 \sqrt{\frac{g_{\sim\epsilon}\dot{q}^\epsilon\dot{q}^\sim}{c^2}}\sqrt{1-\frac{V_{qu(n)}}{mc^2}}\left(u_\sim u_\epsilon - g_{rs}u^r u^s g_{\sim\epsilon}\right). \qquad (55)$$

And where the cosmological energy-impulse density $\Lambda$ [68], for eigenstates, reads

$$\Lambda = \frac{8f\,G}{c^4}/\!Œ_n/^2 \quad L_0 = \frac{8f\,G}{c^4}\frac{m/\!Œ_n/^2\,c^2}{\chi} \tag{56}$$

where for scalar uncharged particles leads to

$$L_0 = \lim_{\hbar \to 0} L = (\pm) - \frac{mc^2}{\chi}. \tag{57}$$

The quantum energy impulse tensor density for the particle density distribution $/\!Œ_n/^2$ owing with the hydrodynamic impulse (6), as a function of the wave function reads

$$T_{\sim\epsilon} = T_\sim{}^r g_{r\epsilon} = /\!Œ/^2\, m^2 c^4 \left(\frac{\partial S}{\partial t}\right)^{-1}\left(\frac{p_\sim p_\epsilon}{m^2 c^2} - \left(1 - \frac{V_{qu}}{mc^2}\right)g_{\sim\epsilon}\right)$$

$$= m/\!Œ/^2\, c^2 \left[\frac{\frac{\hbar}{2im^2c^2}\,\partial \ln[\frac{Œ}{Œ*}]}{\partial t}\right]^{-1}$$

$$\left(\left(\frac{\hbar}{2mc}\right)^2 \frac{\partial \ln[\frac{Œ}{Œ*}]}{\partial q^\sim}\frac{\partial \ln[\frac{Œ}{Œ*}]}{\partial q^\epsilon} + \left(1 - \frac{V_{qu}}{mc^2}\right)g_{\sim\epsilon}\right) \tag{58}$$

*Charged Boson in Non-Euclidean Space–Time*

The KGE in non-Euclidean space–time for electromagnetic charged boson

$$\frac{1}{\sqrt{-g}} D^\sim \left(\sqrt{-g}\, D_\sim Œ\right) - \frac{A_\sim \partial^\sim \sqrt{-g}}{\sqrt{-g}} = -Œ\left(\frac{m^2 c^2}{\hbar^2} - \}\,/\!Œ/^2\right) \tag{59}$$

leads to the hydrodynamic system of equations

$$\frac{1}{\sqrt{-g}}\partial_\sim \sqrt{-g}\left(g^{\sim\epsilon}/\!Œ/^2\left(\partial_\epsilon + eA_\epsilon\right)\right) = 0 \tag{60}$$

$$g^{\sim\epsilon}\left(\partial_\sim S + eA_\sim\right)\left(\partial_\epsilon S + eA_\epsilon\right) = m^2 c^2 - mV_{qu} \tag{61}$$

where

$$V_{qu} = -\frac{\hbar^2}{m}\left(\frac{1}{/\!Œ/\sqrt{-g}}\partial_\sim \sqrt{-g}\left(g^{\sim\epsilon}\partial_\epsilon /\!Œ/\right) - \}\,/\!Œ/^2\right), \tag{62}$$

Moreover, the Lagrangean motion equations read

$$p_{n_\sim} = -\frac{\partial L_n}{\partial \dot{q}^\sim}$$

$$D_t p_{n_\sim} = -\frac{\partial L_n}{\partial q^\sim}, \tag{63}$$

where

$$L_n = \left( -mc^2 \sqrt{\frac{g_{\sim\epsilon} \dot{q}^\epsilon \dot{q}^\sim}{c^2}} \sqrt{1 + \frac{\hbar^2}{m^2 c^2} \left( \frac{\partial_\sim \partial^\sim |Œ_n|}{|Œ_n|} - \} |Œ_n|^2 \right)} - g_{\sim\epsilon} e A^\epsilon \dot{q}^\sim \right), \qquad (64)$$

and to the QIETD

$$T_{n\sim}{}^\epsilon = -m |Œ_n|^2 c^2 \sqrt{\frac{g_{\sim\epsilon} \dot{q}^\epsilon \dot{q}^\sim}{c^2}} \sqrt{1 - \frac{V_{qu(n)}}{mc^2}}$$

$$\left( \left( u_\sim + \left( \sqrt{1 - \frac{V_{qu(n)}}{mc^2}} \right)^{-1} \frac{e}{mc} A_\sim \right) u^\epsilon \right.$$

$$\left. - \left( 1 + \left( \sqrt{1 - \frac{V_{qu(n)}}{mc^2}} \right)^{-1} \frac{e}{mc} g_{rs} A^s u^r \right) u_\sim{}^\epsilon \right) \qquad (65)$$

## 4. Discussion and Conclusions

Even if the GE for quantum systems (54) is formally equal to the Einstein GE, it actually demonstrates some differences. If the Einstein GE refers to the gravitational description of classical masses with no information, either about the quantum mechanical properties of bodies or about the string properties of elementary particles building up the matter, the GE (54) introduces the quantum properties in the gravity of a mass distribution that leads to a cosmological term, that is not a simple constant, but rather, has the form of an energy–impulse tensor density that leads to the correct value of the cosmological constant on the galactic space for the description of the motion of galaxies [68].

Even if the gravity of a quantum mechanical system refers to non-quantized fields, it is not without interest, since it is able to define the gravitational coupling of classical fields and their commutation rules in the non Euclidean space–time for the derivation of the quantum gravity equation of quantum fields.

The GE (54) together with (53) allows also to derive the mass eigenvalues of a Schwarzschild black hole [45,68] (like the e-atom model that is currently of interest to the scientific community [38] (and references therein)).

Finally, in this work the hydrodynamic model for bosons described by a KGE with perturbative self-interacting field is derived for defining the coupling with the Einstein gravitational equation of quantum mechanical systems.

The hydrodynamic treatment has a classical-like structure where the quantum dynamics (quantization) is introduced by the quantum potential through the factor $\sqrt{1 - \frac{V_{qu(n)}}{mc^2}}$.

This can be clearly seen by the Lagrangean function $L_n = -\frac{mc^2}{x} \sqrt{1 - \frac{V_{qu(n)}}{mc^2}} = -mc^2 \sqrt{\frac{g_{\sim\epsilon} \dot{q}^\epsilon \dot{q}^\sim}{c^2}} \sqrt{1 - \frac{V_{qu(n)}}{mc^2}}$ that, in the classical limit for $\hbar \to 0$ (i.e., $V_{qu(n)} = 0$ and $\sqrt{1 - \frac{V_{qu(n)}}{mc^2}} = 1$), leads to the known classical expression.

Given the classical-like hydrodynamic model of quantum mechanics, the energy–impulse tensor for the Einstein gravitational equation of quantum mechanical systems is straightforwardly derived. The availability of such a description for mesons facilitates the coupling of them to the GE.

The biunique correspondence between the standard quantum mechanics and the hydrodynamic representation [60–63,70,71,77] warrants that the quantum energy–impulse tensor density is independent by the used formalism.

**Conflicts of Interest:** The author declares no conflict of interest.